\documentclass[twocolumn,amsmath,amssymb]{revtex4}

\usepackage{graphicx}
\usepackage{amsmath}
\usepackage{epsfig}
\usepackage{dcolumn}
\usepackage{bm}

\begin{document}
\begin{abstract}
Here we build some effective boundary conditions to be used in numerical
calculations in order to avoid the thin meshing usually required in problems
involving Hartmann layers near a locally plane wall. Wall model are provided
for both tangential and normal electric current density and velocity. In
particular, a condition on the normal derivative of the tangential velocity
is derived. A wide variety of problems is covered as the only restriction is
that the magnetic Reynolds number has to be large at the scale of the
Hartmann layer. The cases of perfectly conducting or insulating wall are
examined, as well as the case of a thin conducting wall. The newest result
is a condition on the normal velocity accounting for inertial effects in the
Hartmann layer.

\end{abstract}


\title{Effective boundary conditions for magnetohydrodynamic flows with thin
Hartmann layers.}
\author{A. Poth\'{e}rat$^{1}$}
\author{J. Sommeria$^{2}$}
\author{R. Moreau$^{1}$}
\affiliation{$^{1}$Laboratoire EPM-MADYLAM (CNRS),
ENSHMG BP 95 38402 Saint Martin d'H\`{e}res Cedex.}
\affiliation{$^{2}$Laboratoire de Physique (CNRS),
\'Ecole Normale Sup\'{e}rieure de Lyon, 
46 all\'{e}e de l'Italie 69364 Lyon Cedex 07}
\email{e-mail : ap312@eng.cam.ac.uk}
\date{8 October, 2001}

\maketitle

\section{Introduction}

The flow of an electro-conducting fluid near a wall transverse to a magnetic
field produces a specific boundary layer, the Hartmann layer, resulting from
the balance between the Lorentz and the viscous force. This layer is
generally very thin, which is problematic for a direct numerical resolution
of the magnetohydrodynamic equations. The Hartmann layer thickness scales as
the inverse of the magnetic field component orthogonal to the wall. Strong
magnetic fields therefore dramatically reduce the layer thickness, to
typically a few hundredths of \textit{mm} in a liquid metal in a magnetic field of
one Tesla, so that a very fine numerical mesh would be required for a direct
computation. An insufficient resolution could spoil the whole computation as
in some cases, the outer velocity is controlled by the total electric
current passing through the layer, so an accurate description of the
Hartmann layer is essential.

A convective flow in a rectangular elongated cavity alternatively has been
modeled either meshing the layer or using a
simple analytical model (using the classical linear model
simply built on the balance between Lorentz and viscous forces (see for
instance \cite{moreau90}). It was found that 15 meshes within the layer were
required in order to have a discrepancy smaller than $10\%$ from the outer
velocity computed from the analytical model. M\"{u}ck \textit{et. al.} \cite{muck00}
have also performed accurate numerical simulations using such a simple
model. Similar difficulties would arise in many examples such as lithium
blankets designed for nuclear fusion reactors (\cite{buhl96}), where a
liquid metal must evacuate heat under the strong transverse magnetic field
used to confine the plasma, or in electromagnetic steerers where liquid
steel is driven by a sliding magnetic field between transverse plates. Note
finally the applications to the Earth liquid metal core, where we expect
Hartmann layers with thickness less than one meter to occur at the contact
with the mantle or with the solid metal inner core.

We propose here as an alternative approach to use an analytical model of the
Hartmann layer, and to deduce effective boundary conditions for the core
flow. Many previous works use a similar asymptotic approach, leading to a
coupled analytical description of the core flow and boundary layers. Such
fully analytical solutions are however generally limited to linear problems,
\textit{e.g.} Hunt and Shercliff \cite{hunt71}  for duct flows, Walker \cite{walker81} 
for convection. A two-dimensional evolution equation for a 2D core flow relying 
on a similar idea  is proposed in \cite{sm82}. Poth\'{e}rat \textit{
et Al.} \cite{psm00}  extended such effective 2D models to the case of moderate
magnetic fields, taking into account recirculating secondary flows in the
Hartmann layer. Here, we consider again the same effects, but with the goal
of extracting an effective boundary condition for the core flow, without any
assumption on its dynamics (it is not necessarily described as a
two-dimensional flow).

Our approach here is quite general, with possibly non-uniform or time
varying magnetic fields and various wall electric conditions, as presented
in section {\bf 2}. We use a systematic expansion for the Hartmann layer valid for
large magnetic fields, and the matching with the core provides the requested
effective boundary conditions. The zero order classical Hartmann layer is
recalled in section {\bf 3}, and effective boundary conditions are deduced. It is
shown that in many cases, one can just forget the Hartmann layer and allow
the fluid to slip on the wall.

However this is not always sufficient, in particular in the case of
insulating walls. Thus the electric current sheet generated in the Hartmann
layer transmits a friction effect in the core. This is taken into account by
an effective condition on the normal current density. Similarly a normal
velocity is generated in the Hartmann layer, due to the recirculating flows
induced by inertial effects. This appears as a higher order correction on
the Hartmann layer, derived in section {\bf 4}. In the case of Coriolis effects,
these effects are modified as the Hartmann layer is transformed into
Hartmann-Ekman layer, as discussed in section {\bf 5}.

\section{Description of the system}

Let us consider an incompressible fluid with density $\rho$, kinematic
viscosity $\nu $, conductivity $\sigma $ in a magnetic field $\mathbf{B}%
\left( \mathbf{x,} t\right) $ possibly depending on time $t$ and position
vector $\mathbf{x}$. All variables are non-dimensional, using a typical
magnetic field value $B_0$, a velocity $U$, a scale $a$ of the fluid domain.
The time is normalized by the advective time scale $a/U$. The dynamics then
depend on three non-dimensional parameters, the Hartmann number $Ha$, the
interaction parameter $N$ and the magnetic Reynolds number $Rm$, 
\begin{equation}
Ha=(\tfrac{\sigma B_0^2}{\rho}\tfrac{a^2}{\nu})^{1/2}\;\;,\;\;N=\tfrac{%
\sigma B_0^2}{\rho}\tfrac{a}{U} \;\;,\;\;Rm=\mu \sigma U a.
\end{equation}
The Hartmann number and interaction parameter compare the electromagnetic
forces respectively to viscosity and inertia. Note that the hydrodynamic
Reynolds number $Re$ can be expressed as $Re=Ha^2/N$, while the magnetic
Reynolds number $Rm$ compares advection and diffusion of the magnetic field.

The non-dimensional velocity field $\mathbf{u}$ satisfies the Navier-Stokes
equations 
\begin{equation}
\partial _{t}\mathbf{u+u.\nabla }\mathbf{u}={\ \tfrac{1}{Re}}\mathbf{\Delta u%
}+N\mathbf{j\times B}-N{\nabla p},  \label{navier}
\end{equation}

\begin{equation}
\mathbf{\nabla .u}=0  \label{div u},
\end{equation}

with an electromagnetic force proportional to the current density $\mathbf{j}
$, normalized by its estimate $\sigma UB_{0}$. The pressure $p$ has been
normalized by its estimate $N\rho U_{0}^{2}$, corresponding to a balance
between pressure and electromagnetic forces. The electric current and
magnetic field satisfy the Ohm's law and the equations of magnetic
induction, in non-dimensional form, 

\begin{equation}
\mathbf{j}=\mathbf{E}+\mathbf{u\times B},  \label{ohm}
\end{equation}

\begin{equation}
\nabla \times \mathbf{E}=-\partial _{t}\mathbf{B},  \label{rotE}
\end{equation}

\begin{equation}
\mathbf{\nabla \times B=}Rm\,\mathbf{j},  \label{maxwell}
\end{equation}

\begin{equation}
\mathbf{\nabla .B}=0.  \label{div B}
\end{equation}

Note that taking the curl of the Ohm's law (\ref{ohm}) and using (\ref{rotE}%
), we get : 

\begin{equation}
\nabla \times \mathbf{j}=-\partial _{t}\mathbf{B}-\mathbf{u.\nabla B}+%
\mathbf{B.\nabla u}.  \label{curlohm}
\end{equation}

The boundary condition for the velocity is the classical no-slip condition
 
\begin{equation}
\mathbf{u}=0.  \label{uwall}
\end{equation}

For the magnetic field, there is a condition of continuity with the outside
of the normal component $\mathbf{B.n}$ ($\mathbf{n}$ is the normal to the
wall). There is also a condition on the tangential components of $\mathbf{B}$%
, related to the electrical boundary conditions by the induction law (\ref%
{maxwell}). Denoting $\mathbf{j_{W}}$ the current density at the wall, the
tangential projection of the magnetic field is determined from the normal
current $\mathbf{j_{W}.n}$ on the whole boundary, while the normal
derivative of this tangential magnetic field is proportional to the
tangential projection of $\mathbf{j_{W}}$. We shall consider four cases:

\begin{enumerate}
\item  insulating walls, $\mathbf{j_W.n}$=0,
\item  electrodes controlling the injected current $\mathbf{j_W.n}=j_I$ is given,
\item  perfectly conducting walls $\mathbf{j_{W \Vert}}=0$,
\item  thin conducting walls, with conductance $\Sigma _{W}\sigma a$ (so that $%
\Sigma _{W}$ is non-dimensional). Then the tangential electric field $%
\mathbf{E_{\Vert }}$, which is continuous at the boundary, is proportional
to the surface current density $\mathbf{J_{W}}$ along the wall shell, $%
\mathbf{E_{\Vert }=J_{W}}/\Sigma _{W}$. The current conservation in this
shell yields $\nabla _{\Vert }.\mathbf{J_{W}}=j_{I}-j_{Wz}$, where $j_{I}$
is a current density possibly injected on the shell by electrodes from
outside. Then, using the Ohm's law (\ref{ohm}) at the wall, with $\mathbf{u}%
=0$, we get the electric boundary condition
 
\begin{equation}
\nabla _{\Vert }.(\Sigma _{W}\mathbf{j_{W\Vert }})+\mathbf{j_{W}.n}=j_{I},
\label{elbound}
\end{equation}

which in fact covers the four cases. The case of insulating walls or imposed
normal current is obtained with $\Sigma _{W}=0$ and the case of perfectly
conducting walls with $\Sigma _{W}\rightarrow \infty $.
\end{enumerate}

Near the walls with non-zero transverse magnetic field $\mathbf{B.n}$, a
Hartmann boundary layer occurs. It is dominated by a balance between the
electromagnetic force, pressure force and viscosity (the three terms in the
right hand side of (\ref{navier})). The thickness of this layer is in $a(Ha%
\mathbf{B.n})^{-1}$, which we suppose to be much smaller than $a$. This is
verified in most cases of interest. We furthermore assume that the magnetic
field variation $\delta \mathbf{B}$ across this layer is small, $\delta 
\mathbf{B}\ll \mathbf{B}$. Then the magnetic field can be assumed given when
the dynamics of the boundary layer is studied. This is satisfied when the
magnetic Reynolds number at the scale of the Hartmann number is small, $%
Rm/Ha\ll 1$, a condition which is in practice always verified, even if $Rm$
is large. In some engineering applications (e.g. in induction pumps), a
magnetic field oscillation is externally imposed with frequency $f$. Then
our analysis applies if the skin depth $(\mu \sigma f)^{-1/2}$ remains
larger than the Hartmann layer thickness $a(Ha\mathbf{B.n})^{-1}$, so that
again the magnetic field can be considered as uniform across the Hartmann layer.

Inertial effects are assumed small in the Hartmann layer, which is satisfied
for high interaction parameters $N$. We shall consider the perturbative
effects of inertia, resulting in recirculation effects, so our analysis
extends in reality for values of $N$ close to unity. In summary our analysis
applies when 

\begin{equation}
Ha\;\mathbf{B.n}\gg 1,\quad N \gg 1,\quad \frac{Rm}{Ha}\ll 1,\quad f \ll 
\dfrac{Ha^{2}}{\mu \sigma a^{2}}.  \label{conditions numbers}
\end{equation}

Furthermore we shall assume that the curvature radius of the walls is large
with respect to the thickness $aB_0(Ha\mathbf{B.n})^{-1}$ of the Hartmann
layer, so that the latter can be assumed locally plane.

In addition we shall separately discuss the case with strong Coriolis force,
as relevant in a planetary liquid metal core. Then Hartmann-Ekman layers are
obtained instead of Hartmann layers.

\section{The Hartmann Layer.}

\subsection{The equation of motion in the Hartmann layer.}

\begin{figure*}
\centering
\includegraphics[width=0.7\textwidth]{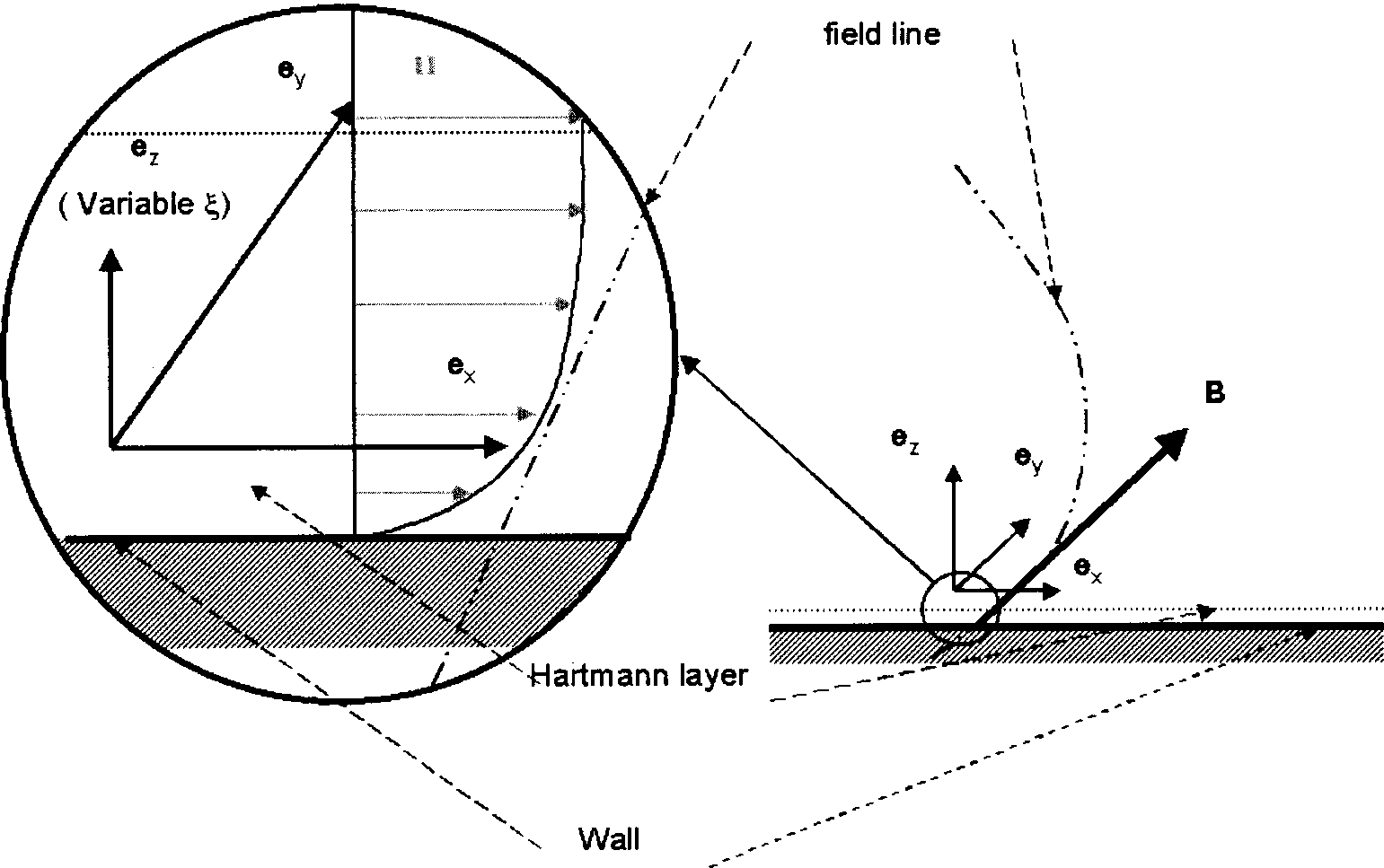}
\caption{Geometric configuration}
\label{config}
\end{figure*}

In the Hartmann boundary layer, the normal derivative dominates the
tangential ones. We choose an orthonormal reference frame ($\mathbf{e}_{x},%
\mathbf{e}_{y},\mathbf{e}_{z}$), where $\mathbf{e_{z}}$ is the unit vector
normal to the wall, and the origin is chosen so that the wall corresponds to
the surface $z=0$. We define the stretched non-dimensional coordinate $\xi
=Ha\;z $, which remains of order one in the Hartmann layer, and denote the
fields by the superscript $\hat{()}$, to specify that they are functions of
this stretched coordinate.

We assume that the wall curvature is sufficiently small, so that the
Hartmann layer can be calculated with the cartesian coordinates $(x,y,\xi )$%
, where the origin of $\xi $ is at the wall. In further discussions, vectors
belonging to the plane of the wall are referred as tangential whereas
vectors orthogonal to the wall are called normal

The velocity $%
\mathbf{u}$ is then decomposed in its tangential projection, denoted $%
\mathbf{{u}_{\Vert }}$, and normal component, denoted ${w}$.

With these conventions, the continuity equation (\ref{div u}) rewrites
 
\begin{equation}
-\dfrac{1}{Ha}\mathbf{\nabla }_{\Vert }.\mathbf{\hat{u}_{\Vert }}=\partial
_{\xi }\hat{w},  \label{cont}
\end{equation}

so that the normal velocity $\hat{w}$ is of order $\frac{1}{Ha}$

The tangential projection of the Navier-Stokes equation yields
 
\begin{eqnarray}
\tfrac{1}{N}\left[ \left( \partial _{t}+\mathbf{\hat{u}.\nabla }_{\Vert
}+Ha\,\hat{w}\,\partial _{\xi }\right) \mathbf{\hat{u}_{\Vert }}\right] -%
\tfrac{1}{Ha^{2}}\mathbf{\Delta }_{\Vert }\mathbf{\hat{u}} \nonumber \\
=-\mathbf{\nabla }%
_{\Vert }\hat{p}+\partial _{\xi \xi }^{2}\mathbf{\hat{u}_{\Vert }}+B_{z}%
\mathbf{\hat{\jmath}\times e}_{z}+\hat{\jmath}_{z}\mathbf{e}_{z}\times 
\mathbf{B}_{\Vert }.  \label{NS}
\end{eqnarray}

The normal component of the Navier-Stokes equation yields, to an excellent
approximation a balance between normal pressure and electromagnetic force,
namely 

\begin{equation}
\mathcal{O}(N^{-1}Ha^{-1})=-\partial_z p +\mathbf{\hat{\jmath}_\Vert\times
B_\Vert }.  \label{NSz}
\end{equation}

The electric current conservation writes :
 
\begin{equation}
-\dfrac{1}{Ha}\mathbf{\nabla }_{\Vert }\mathbf{.\hat{\jmath}_{\Vert }}%
=\partial _{\xi }\hat{\jmath}_{z}.  \label{Current conservation}
\end{equation}

The curl of the Ohm's law (\ref{curlohm}) is written in terms of the
stretched variable as 

\begin{eqnarray}
\dfrac{1}{Ha}\left[ \left( \mathbf{e}_{z}\times \mathbf{\nabla }_{\Vert
}\right) \hat{\jmath}_{z}-\dfrac{d}{dt}\mathbf{B}_{\Vert }+\left( \mathbf{B}%
_{\Vert }\mathbf{.\nabla }\right) \mathbf{\hat{u}_{\Vert }}\right] \nonumber \\ 
=\mathbf{e}_{z}\times \partial _{\xi }\mathbf{\hat{\jmath}_{\Vert }}-B_{z}\partial
_{\xi }\mathbf{\hat{u}_{\Vert }},  \label{curl of Ohm}
\end{eqnarray}

where $\frac{d}{dt}\equiv \partial _{t}+\mathbf{\hat{u}.\nabla }_{\Vert }$
is the advection operator.

The Hartmann layer solution is supposed to match the core solution of the
motion equation. The latter differs from the Hartmann layer solution by its
typical normal lenghscale which is $\mathcal{O}\left( 1\right) $ and to
which the normal coordinate $z$ normalized by $a$ is associated. Then, the
core solution does not satisfy the boundary condition at the wall. Moreover,
the validity domain of the boundary layer solution does not extend to the
core, so that the matching of the two solutions has to occur at an
intermediate scale $\zeta \left( Ha^{-1}\right) $, possibly depending on $Ha$%
,  satisfying \cite{Kaplun54} : 

\begin{equation}
Ha^{-1}\ll \zeta \left( Ha^{-1}\right) \ll 1.  \label{condition on eta}
\end{equation}

We shall here denote the functions of the core coordinate $z$ by the superscript 
$\check{()}$, to distinguish them from the functions of the stretched
coordinate $\xi$. Matching the core solution to the boundary solution at the
intermediate scale is achieved by the asymptotic condition for any quantity $%
g$ : 

\begin{equation}
\lim_{\substack{ Ha^{-1}\rightarrow 0  \\ z_{\zeta}fixed}}\check{g}\left(
\zeta z_{\zeta }\right) -\hat{g}\left( Ha\zeta z_{\zeta }\right) =0
\label{matching bc.}
\end{equation}

for an appropriate intermediate scale $\zeta(Ha^{-1})$ and for any value of
the argument $z_\zeta$. When the function $\hat{g}$ decays exponentially to
a constant, we can just replace (\ref{matching bc.}) by the simpler
condition 

\begin{equation}
\lim\limits_{_{z\rightarrow 0}}{\check{g}}\left( z\right)
=\lim\limits_{_{\xi \rightarrow \infty }}{\hat{g}}\left( \xi \right)
\label{lim}
\end{equation}

However, we shall find that some perturbative terms, of order $N^{-1}$ or $%
Ha^{-1}$ for instance, increase with $\xi$, so that $\lim\limits_{_{\xi
\rightarrow \infty }}{\hat{g}}\left( \xi \right)=\infty$ and the more
general condition (\ref{matching bc.}) must be used instead of (\ref{lim}).

\subsection{The classical Hartmann layer}

The zero order equations are given by neglecting the terms in $1/N$ and $%
1/Ha $ in (\ref{cont}), (\ref{NS}), (\ref{Current conservation}), (\ref{curl
of Ohm}): we only keep the right-hand terms, of order unity. Taking into
account the no slip condition at the wall, (\ref{cont}) and (\ref{Current
conservation}) yield : 

\begin{equation}
\begin{array}{cc}
\hat{w}=0, & \hat{j_{z}} =j_{Wz}.%
\end{array}
\label{wj(0)}
\end{equation}

By continuity with the core, this gives the effective conditions $\check{w}%
(0)=0$ and $\check{j_z}(0)=j_Wz0$, which just reproduce the wall conditions,
as expected across the thin Hartmann layer.

Using (\ref{wj(0)}), (\ref{NS}) and (\ref{curl of Ohm}) are simplified.
Integrating (\ref{curl of Ohm}) with condition $\mathbf{\hat{\jmath}_{\Vert }%
}(0)=\mathbf{j_{W\Vert }}$ yields a relation between the tangential current
density and velocity 

\begin{equation}
\mathbf{\hat{\jmath}_{\Vert }}(\xi )=\mathbf{j}_{W\Vert }-B_{z}\left( 
\mathbf{e}_{z}\times \mathbf{\hat{u}_{\Vert }}\right),
\end{equation}

which just expresses the Ohm's law (\ref{ohm}) with a constant tangential electric
field across the Hartmann layer. Eliminating the current density
with (\ref{NS}) yields an equation for $\mathbf{\hat{u}_{\Vert }}\left( \xi
\right).$ The latter can be solved using the no-slip condition (\ref{uwall}) 
for the tangential velocity and the matching condition (\ref{matching bc.}),
which at zero order simplifies in $\lim\limits_{_{z\rightarrow 0}}\mathbf{%
\check{u}}\left( z\right) =\lim\limits_{_{\xi \rightarrow \infty }}\mathbf{%
\hat{u}}\left( \xi \right) $, so that finally we get the classical Hartmann
velocity and current density Hartmann profiles, 

\begin{subequations}
\begin{gather}
\mathbf{\hat{u}_{\Vert }}\left( \xi \right) = \mathbf{\check{u}}_{\Vert }(0)%
\left[ 1-e^{-B_{z}\xi }\right],  \label{U(0)} \\
\mathbf{\hat{\jmath}}_{\Vert }\left( \xi \right) = \mathbf{j}_{W\Vert }+B_{z}%
\mathbf{\check{u}_{\Vert }}\left( 0\right) \times \mathbf{e}_{z}\left[
1-e^{-B_{z}\xi }\right] .  \label{j(0)}
\end{gather}
\end{subequations}

The matching with the core yields the conditions :

\begin{eqnarray}
\mathbf{\check{u}_\Vert }\left( 0\right) &=&B_{z}^{-1}\mathbf{j}_{W\Vert }%
\mathbf{\times e}_{z}-B_{z}^{-2}j_{Wz}\mathbf{B}_{\Vert }\times \mathbf{e}%
_{z}- B_{z}^{-2}\mathbf{\nabla }_{\Vert } p,  \label{u(0)-j(0)}\nonumber \\
\\
\mathbf{\check{j}_{\Vert }}\left( 0\right) &=&\mathbf{j}_{W\Vert }+B_{z}%
\mathbf{\check{u}_{\Vert }}\left( 0\right) \times \mathbf{e}_{z}.
\label{u0-j0 ohm}
\end{eqnarray}

By eliminating $\mathbf{j}_{W\Vert }$ in these two relations, and using the
matching $\check{j_{z}}(0)=j_{Wz}$, we get : 

\begin{equation}
{B_{z}}\mathbf{\check{j}_{\Vert }}(0)-\mathbf{B_{\Vert }}(0)\check{j_{z}}%
(0)-B_{z}^{-1}\mathbf{\nabla }_{\Vert }p\times \mathbf{e}_{z}=\mathcal{O}%
\left( N^{-1}\right).  \label{j0-p0}
\end{equation}

This relation just expresses the balance, in tangential projection, between
the electromagnetic force $\mathbf{j\times B}$ and the pressure force.

\subsection{Effective boundary conditions for the core}

We get effective boundary conditions for the core at $z \to 0$ by matching
these results on Hartmann layer for $\xi \to \infty$, or more precisely
using the condition (\ref{matching bc.}).

First, the impermeability condition $w=0$ is just transmitted to the core
boundary thanks to (\ref{wj(0)}). We need two additional effective boundary
conditions, one condition for the current, and one hydrodynamic condition in
order to replace the no-slip wall condition. These are provided by the
continuity of $j_z$ and the two relations (\ref{u0-j0 ohm}) and (\ref{j0-p0}%
).

In the case of a fixed injected current density $j_{Wz}$, for instance with
insulating walls, this condition on $j_{Wz}$ is just transmitted to the core
boundary like the normal velocity. This allows to solve the equations for
the magnetic field and electric current, providing $\mathbf{\check{j}_\Vert}$
near the boundary. Then (\ref{j0-p0}) provides the required hydrodynamic
condition, in terms of pressure. In some cases the typical pressure effects
are of order $N^{-1}$, so this condition is no more effective, and this case
will be discussed below. The additional relation (\ref{u0-j0 ohm}) is not
needed for the effective boundary conditions: it just determines the wall
tangential current density $\mathbf{j}_{W\Vert }$.

In the case of a perfectly conducting wall, $\mathbf{j}_{W\Vert }=0$, then (%
\ref{u0-j0 ohm}) and (\ref{j0-p0}) provide relationships between $\mathbf{%
\check{j}_{\Vert }}(0)$, $\check{j_{z}}(0)$, pressure gradient and velocity.

In the case of a thin wall, combining the divergence of (\ref{u0-j0 ohm})
and (\ref{elbound}) gives :

\begin{equation}
\nabla _{\Vert }.[\Sigma _{W}B_{z}(\mathbf{\check{u}_{\Vert }(}0\mathbf{%
)\times e_{z})}]=(\check{j}_{z}(0)-j_{I})+\nabla _{\Vert }.[\Sigma _{W}%
\mathbf{\check{j}_{\Vert }}(0)],  \label{elec}
\end{equation}

which has to be used in combination with the hydrodynamic condition (\ref%
{j0-p0}).

As already noticed, this condition is only relevant if the pressure effects
are of order unity. When they are not imposed from the outside, like in
ducts, pressure gradients rather tend to scale as $\rho U^{2}$, or $N^{-1}$
in non-dimensional units, so that (\ref{j0-p0}) does not provide any
hydrodynamic condition. It just states that the current is nearly aligned
with the magnetic field, so that the electromagnetic force $\mathbf{j\times B%
}$ is weaker (by a factor at least $N$) than expected from direct
dimensional analysis. The effective hydrodynamic condition can be obtained
in all cases from the curl of the Ohm's law (\ref{curlohm}) at $z=0$ : 

\begin{eqnarray}
\partial _{z}\mathbf{\hat{u}}_{\Vert }-B_{z}^{-1}\left( \mathbf{u}_{\Vert }%
\mathbf{.\nabla }_{\Vert }\right) \mathbf{\check{B}_{\Vert }}%
+B_{z}^{-1}\left( \mathbf{B}_{\Vert }\mathbf{.\nabla }_{\Vert }\right) 
\mathbf{\check{u}_{\Vert }} \nonumber \\
=-B_{z}^{-1}\left[ \left( \mathbf{e}_{z}\times 
\mathbf{\nabla }_{\Vert }\right) \check{j}_{z}-\partial _{t}\mathbf{B}%
_{\Vert }+B_{z}^{-1}\partial _{z}\mathbf{\check{j}}\mathbf{\times e}_{z}%
\right].  \label{effective du wall 1}
\end{eqnarray}

The normal derivative $\partial _{z}\mathbf{\check{j}}\left( 0\right) $ can
be expressed as a function of the current density and the magnetic field by
differentiating the Navier-Stokes equation in the core
at leading order (\ref{u(0)-j(0)}), with respect to $z$. In addition, the pressure can be
eliminated using (\ref{NSz}), the vertical component of the Navier Stokes
equation in the core. Then (\ref{effective du wall 1}) becomes : 

\begin{equation}
\partial _{z}\mathbf{\check{u}}_{\Vert }=-B_{z}^{-1}\left[ 
\begin{array}{c}
\left( \mathbf{e}_{z}\times \mathbf{\nabla }_{\Vert }\right) \check{j}_{z}-%
\frac{d}{dt}\mathbf{B}_{\Vert }\mathbf{+}\left( \mathbf{B}_{\Vert }\mathbf{%
.\nabla }\right) \mathbf{\check{u}} \\ 
+\partial _{z}\mathbf{\check{j}}_{\Vert }\left( 0\right) \times \mathbf{e}%
_{z}%
\end{array}
\right],  \label{effective du wall}
\end{equation}

where $\partial _{z}\mathbf{\check{j}}_{\Vert }\left( 0\right) \times 
\mathbf{e}_{z}$ is obtained from the solution of the system formed with (\ref%
{j0-p0}) and (\ref{NSz}). An alternate way to obtain a condition on $%
\partial _{z}\mathbf{\hat{u}}_{\Vert }(0)$. This relation relates an
effective boundary condition on velocity to the boundary condition for the
electric current. In the case of a magnetic field normal to the wall, $%
\mathbf{B}_{\Vert }=0$, it reduces to 

\begin{equation}
\partial _{z}\mathbf{\check{u}_{\Vert}}^{\left( 0\right) }\left( 0\right)
=B_{z}^{-1}\left( \mathbf{e}_{z}\times \mathbf{\nabla }_{\Vert }\right)
j_{zW}^{\left( 0\right) }  \label{simpdu}
\end{equation}

and $\partial _{z}\mathbf{\hat{u}}_{\Vert }=0$ for an insulating wall. Note
that this effective boundary condition has been already derived by Sommeria
and Moreau (1982) to justify the two-dimensional dynamics of turbulence
observed in duct flows with insulating walls and transverse magnetic field.
When an electric current is injected through electrodes at the boundary, a
normal shear is introduced as indicated by (\ref{simpdu}). This corresponds
to the existence of a shear layer parallel to the magnetic field,
propagating from the electrode along magnetic field lines into the core flow.

Up to now, the effective boundary conditions could be obtained without
explicit calculation of the Hartmann layer. We could have just used the
impermeability condition for velocity and continuity of the normal current
density, while keeping free the tangential projections of the velocity and
current. However the previous results do not account for the phenomenon of
Hartmann friction, which is important for a uniform transverse magnetic
field and insulating walls. This effect is due to the closing in the core of
electric current sheets generated in the Hartmann layer. With our approach
it appears as a next order term in the normal current $j_{z}$, as derived
systematically in Appendix /ref{app:A2}. We can find the result more intuitively by
noticing that the Hartmann layer contains a current sheet with surface
density $\int_{0}^{+\infty }\left[ \mathbf{\hat{\jmath}}_{\Vert }\left( \xi
^{\prime }\right) \mathbf{-\check{j}}_{\Vert }\left( 0\right) \right] d\xi
^{\prime }$. Then the current conservation is accounted by an additional
normal current

\begin{equation}
\check{j}_{z}\left( 0\right) =-\tfrac{1}{Ha}\mathbf{\nabla }_{\Vert
}.\int_{0}^{+\infty }\left[ \mathbf{\hat{\jmath}}\left( \xi ^{\prime
}\right) \mathbf{-\check{j}}\left( 0\right) \right] d\xi ^{\prime },
\label{j sheet}
\end{equation}

which, applied to (\ref{j(0)}), yields : 

\begin{equation}
\check{j}_{z}=j_{zW}+{\tfrac{1}{Ha}}{\nabla }_{\Vert }\times \mathbf{\check{u%
}}(0).  \label{cond on jz (0,1)}
\end{equation}

This corresponds to the well-know result according to which the normal
current induced outside a Hartmann layer is proportional to the vorticity
outside the layer.

\section{Flow rate out of the Hartmann layer}

The effective condition of zero normal velocity is valid only at zero order.
In reality a small normal velocity can be induced by the Hartmann layer and
this may be important for the convective transport of heat or chemicals at
the wall. This normal velocity is obtained from the divergence of the total
tangential flow rate within the Hartmann layer, as for the current (\ref{j
sheet}) 

\begin{equation}
\check{w}\left( 0\right) =-\tfrac{1}{Ha}\mathbf{\nabla }_{\Vert
}.\int_{0}^{+\infty }\left[ \mathbf{\hat{u}}\left( \xi ^{\prime }\right) 
\mathbf{-\check{u}}\left( 0\right) \right] d\xi ^{\prime }.  \label{w sheet}
\end{equation}

Plugging (\ref{U(0)}) into (\ref{w sheet}) yields the normal velocity related
to the classical Hartmann layer profile : 

\begin{equation}
\check{w}\left( 0\right) =\mathbf{\nabla }_{\Vert }.\left( \mathbf{\check{u}}%
\left( 0\right) B_{z}^{-1}\right).  \label{w(0,1)}
\end{equation}

This velocity just represents the flow over a weak topography with height $%
\dfrac{aB_0B_{z}^{-1}}{Ha}$ (in real units), which corresponds to the thickness
of the Hartmann layer. Indeed this is a zone of stagnant fluid and the core
flow has to move around it. When following a fluid particle in its
tangential motion near the wall, this normal motion is reversible and does
not provide normal transport of matter.

The true transport is obtained at next order by perturbing the Hartmann
layer basic profile with terms in $N^{-1}$. These effects are significant in
practice for small hydrodynamic scales (see \cite{psm00})
over which the magnetic field is uniform, which simplifies
calculations. Then the velocity profile in the Hartmann layer is perturbed
by the inertial terms expressed with the basic profile of the tangential
velocity (\ref{U(0)}). At this order, the tangential velocity
profile becomes (see appendix \ref{app:B}) :

\begin{eqnarray}
\mathbf{\hat{u}}_{\Vert }\left( \xi \right) = \nonumber \\
\left[ \mathbf{\check{u}}_{\Vert }\left( 0\right) -B_{z}^{-2} \left( \tfrac{4}{9}B_{z}^{-1}\mathbf{s}%
_{1}-\tfrac{1}{3}\mathbf{s}_{0}\right) e^{-B_z\xi }\right] \left(
1-e^{-B_{z}\xi }\right) \nonumber \\ +B_{z}^{-2}\xi e^{-B_z \xi} 
\left[ \tfrac{1}{3}\mathbf{s}_{1}\mathbf{%
-r}_{0}\tfrac{B_{z}}{2}-\tfrac{1}{4}\mathbf{r}_{1}\left( 1+B_{z}\xi \right) %
\right],  \label{u10} 
\end{eqnarray}

\begin{subequations}
\begin{eqnarray}
\mathbf{\check{u}}_{\Vert }\left( 0\right) &=&B_{z}^{-2}\Big[ -
\mathbf{\nabla }_{\Vert }p+B_{z}\mathbf{j}_{\Vert W}\times \mathbf e_z \nonumber \\
&&-j_W \mathbf{B}_{\Vert }\mathbf{\times e}_{z}-\dfrac{1}{N}\dfrac{d\mathbf{\check{u}%
}_{\Vert}}{dt}\Big]  \label{u(1,0)inf},\nonumber \\ \\
\mathbf{r}_{0} &=&- \partial _{t}\mathbf{\check{u}}_{\Vert }-2
\left( \mathbf{\check{u}}_{\Vert }\mathbf{.\nabla_{\Vert} }\right) \mathbf{\check{u}}%
_{\Vert }, \\
\mathbf{r}_{1} &=&- \left[ \mathbf{\check{u}}_{\Vert }\mathbf{.\nabla 
}_{\Vert }B_{z}\right] \mathbf{\check{u}}_{\Vert }+ \mathbf{\check{u}}%
_{\Vert }\partial _{t}B_{z}, \\
\mathbf{s}_{0} &=& \left( \mathbf{\check{u}}_{\Vert }\mathbf{.\nabla }_{\Vert}%
\right) \mathbf{\check{u}}_{\Vert }, \\
\mathbf{s}_{1} &=&- \left[ \mathbf{\check{u}}_{\Vert }\mathbf{.\nabla 
}_{\Vert }B_{z}\right] \mathbf{\check{u}}_{\Vert }.
\end{eqnarray}
\end{subequations}

The condition (\ref{u(1,0)inf}) replaces (\ref{u(0)-j(0)}).

The corresponding normal velocity is obtained by the mass conservation law (%
\ref{w sheet}), which yields

\begin{eqnarray}
\check{w}\left( 0\right) =B_{z}^{-1}\mathbf{\nabla }_{\Vert }\mathbf{\check{u%
}}\left( 0\right) +\dfrac{B_{z}^{-4}}{HaN}\mathbf{\nabla }_{\Vert } \nonumber\\
.\left[ 
\tfrac{B_{z}}{2}\partial _{t}\mathbf{\check{u}}\left( 0\right) -\tfrac{5}{6}%
\left( \mathbf{\check{u}.\nabla }_{\Vert }\right) \mathbf{\check{u}}-\tfrac{%
B_{z}}{6}\mathbf{\check{u}}\partial _{t}B_{z}\right].  \label{effective w 1/N}
\end{eqnarray}

In the case of a steady uniform magnetic field the expression of the normal
velocity simplifies : 

\begin{eqnarray}
\check{w}\left( 0\right) =-Ha^{-1}N^{-1}B_{z}^{-2}\mathbf{\nabla }_{\Vert }.%
\left[ \tfrac{5B_{z}}{6}\left[ \mathbf{\check{u}}.\mathbf{\nabla }_{\Vert }%
\right] \mathbf{\check{u}}\right] \nonumber \\
+\mathcal{O}_{i+j=3}\left(
N^{-i}Ha^{-j}\right).  \label{inertial recirculation}
\end{eqnarray}

This expression is almost the same as the one found in \cite{psm00} 
in the case of a flow between two transverse plates. However,
in this latter configuration, the quasi two dimensionality of the core
allows to consider $\mathbf{\nabla }_{\Vert }.\mathbf{\check{u}}^{\left(
0\right) }\simeq 0$ at the leading order. Then, the vertical velocity at the
edge of the Hartmann layer is exclusively the consequence of inertial
effects arising in the Hartmann layer. If the flow is axisymetric as under a
big vortex, (\ref{inertial recirculation}) simply expresses the secondary
flow due to the Ekman recirculation.

\section{The Hartmann-Ekman layers.}

If the motion is described in a frame of reference which is in rotation
around an axis perpendicular to the wall (speed $\Omega $), a Corolis force
appears in the right hand side of (\ref{NS}). We shall write it $A^{-1}%
\mathbf{u}_{\Vert }\mathbf{\times e}_{z}$ using non-dimensional coordinates,
where $A=\frac{\sigma B^{2}}{2\rho \Omega }$ is the Elsasser number.
Assuming that the magnetic field is permanent and orthogonal to the wall ($%
B_{z}=1$) and neglecting the other inertial terms allows to find an
expression for the vertical velocity as in \cite{Acheson73}:

\begin{eqnarray}
\mathbf{\hat{u}}_{\mathcal{\Vert }}&(\xi ) &= \nonumber \\
&&\mathbf{\check{u}}_{\mathcal{\Vert }}(z=0)(1-e^{-c\xi }\cos b\xi )\nonumber \\ 
&-&\frac{1}{bc}[ A^{-1}\mathbf{e}_{z}\times \mathbf{\hat{u}}_{\mathcal{\Vert }}(z=0) \nonumber \\
&+& (1+c^{2}-b^{2}) \mathbf{\hat{u}}_{\mathcal{\Vert }}(z=0)]%
e^{-c\xi }\sin b\xi  \label{u Ha-Ek} \\
\mathbf{\check{u}_{\perp }}&\left( z=0\right)& = \nonumber \\
&\frac{1}{1+A^{-2}}&\left[\left[ \mathbf{\mathbf j}_{W\Vert }-A^{-1}\mathbf{\nabla }%
_{\Vert }\hat{p}\right] \times \mathbf{e}_{z}-\mathbf{\mathbf j}_{W \Vert
} -A^{-1}\mathbf{\nabla }_{\Vert }\hat{p}\right] \nonumber \\
\label{u-j HaEk}
\end{eqnarray}

where $c=-\left( 1+A^{-2}\right) ^{\frac{1}{4}}\cos \left[ \frac{1}{2}%
\arctan A^{-1}\right] $ and $b=-\left( 1+A^{-2}\right) ^{\frac{1}{4}}\sin %
\left[ \frac{1}{2}\arctan A^{-1}\right] $. As the condition (\ref{u0-j0 ohm}%
) is still valid, the discussion on the effective electric boundary
condition still applies, replacing (\ref{u(0)-j(0)}) with (\ref{u-j HaEk}).

The normal velocity associated with this profile has the same expression as
for a classical Ekman layers but, with a thickness modified by the magnetic
field 

\begin{eqnarray}
\hat{w}\left( 0\right) = 
Ha^{-1}[ \left( 1+A^{-2}\right) ^{-\frac{1}{2}}\mathbf{\nabla_{\Vert }\times \check{u}_{\Vert }}\left( z=0\right) 
\nonumber \\
-\mathbf{\nabla }_{\mathbf{\Vert }}\mathbf{.\check{u}_{\Vert }}\left( z=0\right)]   \label{w Ek Ha}
\end{eqnarray}%

This result is not surprising since at the edge of Hartmann layers, the
normal velocity occurs either because of an additional effect (such as
inertial) or because $\mathbf{\nabla }_{\mathbf{\Vert }}\mathbf{.\check{u}%
_{\perp }}\left( z=0\right) \neq 0$. Therefore, the normal velocity in
Hartmann-Ekman layers only arises because of the Ekman spiral term in $%
\mathbf{e}_{z}\times \mathbf{\hat{u}}_{\mathcal{\Vert }}$ as in Ekman layers.

The same remark applies for the electric current density except that it
results from the Hartmann behavior of the layer. 

\begin{eqnarray}
\check{\jmath}_{z}\left( 0\right) =\jmath _{Wz}-Ha^{-1}[ \mathbf{\nabla
_{\Vert }\times \check{u}_{\Vert }}\left( z=0\right) \nonumber \\
-\left( 1+A^{-2}\right)
^{-\frac{1}{2}}\mathbf{\nabla }_{\mathbf{\Vert }}\mathbf{.\check{u}_{\perp }}%
\left( z=0\right) ].  \label{jz Ha-Ek infinity}
\end{eqnarray}

These results are equivalent to (\ref{effective w 1/N}) and (\ref{cond on jz
(0,1)}) with a modification factor for (\ref{w Ek Ha}), due to the fact that
inertia is treated as a perturbation in (\ref{effective w 1/N}).

\section{Conclusion.}

We have obtained effective boundary conditions for the core in the parameter
regime (\ref{conditions numbers}), which is quite commonly reached in
magnetohydrodynamics.

The impermeability condition at the wall is reproduced to a good precision
as an effective condition for the wall. A small normal velocity does however
exist. First, the Hartmann layer is a stagnant zone, and inhomogeneities of
its thickness result in a ``topography'' effect (\ref{w(0,1)}) for the core
flow. More importantly, a pumping flow (\ref{inertial recirculation}) is
driven by weak recirculating flows arising as perturbative effects in the
Hartmann layers. A pumping effect also occurs in Hartmann-Ekman boundary
layers obtained in the presence of Coriolis force.

An important point is that if the magnetic field has a tangential
component or if the wall is conducting, the no-slip condition is replaced by
a condition (\ref{effective du wall}) on the normal shear.

The electric boundary condition provides a normal current (\ref{cond on jz
(0,1)}). The closing of this current in the core is responsible for the
Hartmann friction effects which are important when the tangential electric
current density is weak in the core. The latter case is relevant when the
field is homogeneous and normal to an insulating wall. In this case, the core
flow is quasi 2D. But when the wall is not insulating or when the magnetic
field has a tangential component, strong electric current are passed to the
core and the Hartmann layer is no more active. If the wall is not
insulating, the normal electric current injected in the core has almost the
same value as the electric current at the wall and the condition for the
tangential velocity (\ref{effective du wall}) indicates that the core is
three-dimensional at the edge of the Hartmann layer. The effective
conditions for the electric current are then deduced from (\ref{j0-p0}) and (%
\ref{elbound}). A tangential component of the magnetic field also results in
a non-zero derivative of the tangential velocity in and strong electric
current injected in the core which is also expressed by condition (\ref%
{j0-p0}) and (\ref{u(0)-j(0)}).

\appendix

\section{Appendix: full matching method} 
\label{app:A}

\subsection{expansion in $Ha^{-1}$ and $N^{-1}$}

We are interested in the limit $Ha\gg 1$ and $N\gg 1$ so that each quantity
is developped in terms of these two small parameters : 

\begin{equation}
g=g^{(0)}+g^{\left( 1,0\right) }\dfrac{1}{N}+g^{\left( 0,1\right) }\dfrac{1}{%
Ha}+g^{\left( 1,1\right) }\dfrac{1}{HaN}+...
\label{}
\end{equation}

Following \cite{Cole81}, the matching condition at order $(i,j)$ requires that
there exists $\left( K,L,k,l\right) \in N^{4}$ and an intermediate scale $%
\zeta \left( Ha^{-1}\right) $ satisfying (\ref{condition on eta}) such that:

\begin{widetext}
\begin{multline}
\lim_{\substack{ Ha^{-1}\rightarrow 0  \\ z_{h}fixed}}   
\dfrac{\sum_{n_{1}=0}^{K}\sum_{n_{2}=0}^{L}\check{g}^{\left( n1,n2\right)
}\left( \zeta z_{\zeta }\right)
N^{-n_{1}}Ha^{-n_{2}}-\sum_{n_{1}=0}^{k}\sum_{n_{2}=0}^{l}\hat{g}^{\left(
n1,n2\right) }\left( \zeta z_{\zeta }Ha\right) N^{-n_{1}}Ha^{-n_{2}}}{%
N^{-i}Ha^{-j}}
=0
\label{match_ij}
\end{multline}
\end{widetext}

\subsection{Effective normal velocity and current sheet}
\label{app:A2}

The divergence of the flow sheet $\mathbf{\hat{u}}^{\left( 0\right) }\left(
\xi \right) $ integrated over the Hartmann layer yields the vertical
velocity at the edge of the layer. To demonstrate this result, let's
integrate (\ref{cont})$-Ha^{-1}\times $(\ref{cont})$\left( z=0\right) $
between $0$ and $\zeta \left( Ha^{-1}\right) \,z_{\zeta }\,Ha$ :

\begin{multline}
\hat{w}\left( \xi \right) =\tfrac{1}{Ha}\int_{0}^{\zeta \,z_{\zeta
}\,Ha}\partial _{z}\check{w}\left( 0\right) d\xi ^{\prime } \\
-\tfrac{1}{Ha}%
\int_{0}^{\zeta \,z_{\zeta }\,Ha}\mathbf{\nabla }_{\Vert }.\left[ \mathbf{%
\hat{u}}\left( \xi ^{\prime }\right) \mathbf{-\check{u}}\left( 0\right) %
\right] d\xi ^{\prime }.
\label{cont int -cont}
\end{multline}

Assuming $\zeta $ satisfies (\ref{condition on eta}) and using (\ref{cont
int -cont}) yields : 

\begin{multline}
\lim\limits_{\substack{ \zeta \rightarrow 0  \\ z_{\zeta }\text{fixed}}}%
\hat{w}\left( \xi \right) =\check{w}\left( 0\right) =\lim\limits_{\substack{ %
\zeta \rightarrow 0  \\ z_{\zeta }\text{fixed}}} \\
\left[ \zeta z_{\zeta }\check{%
w}\left( 0\right) 
-\tfrac{1}{Ha}\int_{0}^{\tfrac{\zeta z_{\zeta }}{Ha^{-1}}}%
\mathbf{\nabla }_{\Vert }.\left[ \mathbf{\hat{u}}\left( \xi ^{\prime
}\right) \mathbf{-\check{u}}\left( 0\right) \right] d\xi ^{\prime }\right]
\label{w(0) sheet}.
\end{multline}

If $\mathbf{\hat{u}}\left( \xi ^{\prime }\right) \mathbf{-\check{u}}\left(
0\right) $ is a polynom of exponential function, integral and divergence can
be interverted in the limit $\zeta \rightarrow 0$ . Indeed, although $%
\int_{0}^{\tfrac{\zeta z_{\zeta }}{Ha^{-1}}}\mathbf{\nabla }_{\Vert }.\left[ 
\mathbf{\hat{u}}\left( \xi ^{\prime }\right) \mathbf{-\check{u}}\left(
0\right) \right] d\xi ^{\prime }$ and $\mathbf{\nabla }_{\Vert }.\int_{0}^{%
\tfrac{\zeta z_{\zeta }}{Ha^{-1}}}\left[ \mathbf{\hat{u}}\left( \xi ^{\prime
}\right) \mathbf{-\check{u}}\left( 0\right) \right] d\xi ^{\prime }$ are not
equal to each other, their difference tends toward $0$ in the limit $\zeta
\rightarrow 0$. Finaly, $\check{w}\left( 0\right) $ writes : 

\begin{equation}
\check{w}\left( 0\right) =-\tfrac{1}{Ha}\mathbf{\nabla }_{\Vert
}.\int_{0}^{+\infty }\left[ \mathbf{\hat{u}}\left( \xi ^{\prime }\right) 
\mathbf{-\check{u}}\left( 0\right) \right] d\xi ^{\prime }.
\end{equation}%

We insist that the limit processes have to be carefully applied in (\ref%
{w(0) sheet}) (in particular an asymptotic expansion in terms of $\zeta $
has to be performed on the exponential terms in order to reveal the
different orders in $Ha^{-1}$). These difficulties could be avoided
integrating (\ref{cont}) between $0$ and $\xi $ and applying (\ref{cont int
-cont}) at the considered order but the intuitive result (\ref{w sheet})
would be shortcut.

The same process can be applied to the electric current density and it
yields : 

\begin{equation}
\check{j}_{z}\left( 0\right) =-\tfrac{1}{Ha}\mathbf{\nabla }_{\Vert
}.\int_{0}^{+\infty }\left[ \mathbf{\hat{\jmath}}\left( \xi ^{\prime
}\right) \mathbf{-\check{j}}\left( 0\right) \right] d\xi ^{\prime }
\end{equation}

\section{Appendix: Effective conditions with inertia}
\label{app:B}

We shall now look for the way the effective boundary conditions are affected
by moderate inertial effects. The latter are taken into account by considering $%
N^{-1}\;$order of equations (\ref{NS}) and (\ref{curl of Ohm}). Using
leading order solutions (\ref{U(0)}), (\ref{wj(0)}) and (\ref{j(0)}) to
assess inertial terms, the equation for $\mathbf{\hat{u}}^{\left(
1,0\right) }$ writes 

\begin{multline}
\partial _{\xi \xi }^{2}\mathbf{\hat{u}}^{(1,0)}-B_{z}^{2}\mathbf{\hat{u}}^{(1,0)}=\\
\left[ 
\mathbf{\nabla }_{\Vert }p^{\left( 0\right) }+\dfrac{d\mathbf{\check{u}}%
^{\left( 0\right) }}{dt}-B_{z}\mathbf{j}_{\Vert W}^{\left( 0,1\right)
}\times \mathbf{e}_{z}+j_{W}^{\left( 0,1\right) }\mathbf{B}_{\Vert }\mathbf{%
\times e}_{z}\right] \\
+e^{-B_{z}\xi } \Big[\partial _{t}\mathbf{u}-2\left( \mathbf{\check{u}}%
^{\left( 0\right) }\mathbf{.\nabla }\right) \mathbf{\check{u}}^{\left(
0\right) }+B_{z}^{-2}\left[ \mathbf{\check{u}}^{\left( 0\right) }\mathbf{%
.\nabla }_{\Vert }B_{z}\right] \mathbf{\check{u}}^{\left( 0\right) }\\
-\xi %
\left[ \left[ \mathbf{\nabla }_{\Vert }.\mathbf{\check{u}}^{\left( 0\right) }%
\right] \mathbf{\check{u}}^{\left( 0\right) }-\mathbf{\check{u}}^{\left(
0\right) }\partial _{t}B_{z}\right]\Big] 
+e^{-2B_{z}\xi }\Big[ \left( \mathbf{\check{u}}^{\left( 0\right) }\mathbf{%
.\nabla }\right) \mathbf{\check{u}}^{\left( 0\right) }\\
-B_{z}^{-2}\left[ 
\mathbf{\check{u}}^{\left( 0\right) }\mathbf{.\nabla }_{\Vert }B_{z}\right] 
\mathbf{\check{u}}^{\left( 0\right) }-\xi B_{z}^{-1}\left[ \mathbf{\check{u}}%
^{\left( 0\right) }\mathbf{.\nabla }_{\Vert }B_{z}\right] \mathbf{\check{u}}%
^{\left( 0\right) }\Big] \\
=\left[ \mathbf{d+}\left( \mathbf{r}_{0}+\mathbf{r}_{1}\xi \right)
e^{-B_{z}\xi }+\left( \mathbf{s}_{0}+\mathbf{s}_{1}\xi \right) e^{-2B_{z}\xi
}\right]
\end{multline}

Using the no-slip condition yields : 

\begin{eqnarray}
\mathbf{\hat{u}}(\xi)&=&B_{z}^{-2}\left[ -\mathbf{d}\left( 0\right) +\left( 
\tfrac{4}{9}B_{z}^{-1}\mathbf{s}_{1}-\tfrac{1}{3}\mathbf{s}_{0}\right)
e^{-B_{z}\xi }\right] \left( 1-e^{-B_{z}\xi }\right) \nonumber \\ 
&+&B_{z}^{-2}\left[ 
\tfrac{1}{3}\mathbf{s}_{1}\mathbf{-r}_{0}\tfrac{B_{z}}{2}-\tfrac{1}{4}%
\mathbf{r}_{1}\left( 1+B_{z}\xi \right) \right] \xi e^{-B_{z}\xi }
\end{eqnarray}

Here again, the matching condition (\ref{matching bc.}) simply reduces to $%
\lim\limits_{_{z\rightarrow 0}}\mathbf{\check{u}}\left( z\right)
=\lim\limits_{_{\xi \rightarrow +\infty}}\mathbf{\check{u}}\left( \xi \right)$,
that is : 

\begin{equation}
\mathbf{\check{u}}\left( 0\right) =-B_{z}^{-2}\left[ \mathbf{\nabla }_{\Vert
}p-B_{z}\mathbf{j}_{\Vert W}\times \mathbf{e}_{z}+j_{W}\mathbf{B}_{\Vert }%
\mathbf{\times e}_{z}+\tfrac{1}{N}\dfrac{d\mathbf{\check{u}}^{\left(
0\right) }}{dt}\right].  \label{u(1,0)infB}
\end{equation}

This condition is equivalent to condition (\ref{u(0)-j(0)}). The tangential
current density can be derived from (\ref{curl of Ohm}), neglecting $Ha^{-1}$
terms : 

\begin{equation}
\mathbf{\hat{\jmath}}\left( \xi \right) =-\mathbf{e}_{z}\times \mathbf{%
\hat{u}}\left( \xi \right) +\mathbf{j}_{W}
\end{equation}

Here again, the matching condition can be reduced to a simple limit process
so that it comes finally that (\ref{u0-j0 ohm}) is still valid at order $%
N^{-1}$. The same discussion about the effective electric conditions as in
section \textbf{3.3} also applies.

A vertical velocity order $\mathcal{O}\left( Ha^{-1}N^{-1}\right) $ can be
associated to the inertial jet in the Hartmann layer. But if the field is
not uniform, it is negligible compared to the vertical velocity induced by
non uniformity of the field (\ref{w(0,1)}). Therefore the expression is
computed under the assumption $\partial _{x}B_{z}=\partial _{y}B_{z}=0$.

Using the continuity equation in the layer (\ref{cont}) yields the
expression of the normal velocity within the Hartmann layer 

\begin{multline}
\hat{w}=-B_{z}^{-1}\left[ -1+e^{-B_{z}\xi }+\xi B_{z}\right] \mathbf{\nabla }%
_{\Vert }.\left( \mathbf{d}B_{z}^{-2}\right) \\
+B_{z}^{-2}\left( 1-e^{-B_{z}\xi }-B_{z}\xi e^{-B_{z}\xi }\right) \mathbf{%
\nabla }_{\Vert }.\left( B_{z}^{-2}\left[ \mathbf{r}_{0}\tfrac{B_{z}}{2}+%
\tfrac{1}{4}\mathbf{r}_{1}\right] \right) \\
+\tfrac{B_{z}^{-1}}{2}\left( 1-e^{-B_{z}\xi }\right) ^{2}\mathbf{\nabla }%
_{\Vert }.\left( B_{z}^{-2HaN}\tfrac{1}{3}\mathbf{s}_{0}\right) \\
+\tfrac{B_{z}^{-3}}{12}[ -6+6e^{-B_{z}\xi }+6B_{z}\xi e^{-B_{z}\xi
}+3B_{z}^{2}\xi ^{2}e^{-B_{z}\xi }\\
+B_{z}^{3}\xi ^{3}] \mathbf{\nabla }%
_{\Vert }.\left( B_{z}^{-1}\mathbf{r}_{1}\right).\\
\end{multline}

The velocity in the core is obtained integrating the tangential 
flow rate across the Hartmann layer thanks to (%
\ref{match_ij}) :
\begin{eqnarray}
\check{w}\left( 0\right) &=& -B_z^{-1}\mathbf{\nabla }_{\Vert }.\mathbf{\check{u%
}}\left( 0\right) +\frac{B_{z}^{-4}}{HaN}\mathbf{\nabla }_{\Vert }.%
\Big[ \mathbf{r}_{0}\tfrac{B_{z}}{2}+\tfrac{1}{6}\mathbf{r}_{1} \nonumber \\
&&+\tfrac{B_{z}}{6}\mathbf{s}_{0}\Big]  \nonumber \\
&=&B_{z}^{-4}\mathbf{\nabla }_{\Vert }.\Big[ \tfrac{B_{z}}{2}\partial _{t}%
\mathbf{\check{u}}^{\left( 0\right) }\left( 0\right) -\tfrac{5}{6}\left( 
\mathbf{\check{u}}^{\left( 0\right) }\mathbf{.\nabla }_{\Vert }\right) 
\mathbf{\check{u}}^{\left( 0\right) } \nonumber \\
&&-\tfrac{B_{z}}{6}\mathbf{\check{u}}%
^{\left( 0\right) }\partial _{t}B_{z}\Big].
\label{cond on w(1,1)}
\end{eqnarray}
The latter result is also obtained by direct application of (\ref{w sheet}).

A normal current density also results from the current conservation (\ref%
{Current conservation}), but this vertical current is negligible in front of
(\ref{cond on jz (0,1)}), except in the very particular case of a uniform
field and irrotational outer flow.

\bigskip





%
%



\end{document}